\documentclass[twocolumn,letter]{jpsj2}

\usepackage{amsmath}

\title{
Electric Conductivity of 
the Zero Gap Semiconducting State \\
 in $\alpha$-(BEDT-TTF)$_2$I$_3$  Salt
}

\author{
Shinya Katayama, Akito Kobayashi and Yoshikazu Suzumura
}

\inst{
Department of Physics, Nagoya University, Nagoya 464-8602
}

\recdate{\today}

\abst{
The electric conductivity 
  which reveals the zero gap semiconducting (ZGS) state 
   has been investigated 
    as the function of  temperature $T$ and life time $\tau$ 
  in order to understand the ZGS state in  quarter-filled 
    $\alpha$-(BEDT-TTF)$_2$I$_3$ salt
   with four sites in the unit cell.  
By treating  $\tau$  as a parameter and making use of  
   the one-loop approximation, 
  it is found that the conductivity 
   is proportional to $T$ and $\tau$ for $k_B T\gg\hbar/\tau$ and 
       independent of $T$ and $\tau$ for $k_B T\ll\hbar/\tau$. 
Further   the conductivity being  independent of $T$  in the ZGS state 
 is examined  in terms of Born approximation for the impurity scattering. 
}

\kword{organic conductor, electric conductivity, 
zero gap semiconductor, Dirac cone, multi-band, 
uniaxial pressure, $\alpha$-(BEDT-TTF)$_2$I$_3$
}

\newcommand{\eti}{$\alpha$-(ET)$_2$I$_3$ }

\newcommand{\blet}[1]{\mbox{\boldmath${#1}$}}

\newcommand{\vep}{\varepsilon}
\newcommand{\dsum}[2]{\displaystyle{\sum_{#1}^{#2}}}
\newcommand{\dint}[2]{\displaystyle{\int_{#1}^{#2}}}

\newcommand{\etal}{\textit{et al. }}
\newcommand{\jpsj}[3]{J. Phys. Soc. Jpn. {\textbf{#1}} ({#2}) {#3}}

\begin{document}
\maketitle

 In  quasi-two-dimensional organic conductors, 
  bis(ethylene dithiolo)tetrathiofulvalene (abbreviated as BEDT-TTF) salts,  
    several electronic states, such as 
 Mott inslator, charge ordering and  superconductor, \cite{Ishiguro_1998,Hseo1}
  have been  obtained by varying temperature, 
   hydrostatic pressure and  uniaxial strain. 
Especially,  
  the  unconventional behavior with decreasing temperature
  has been found in  
   the \eti salt ($\alpha$ phase of BEDT-TTF tri-iodide), 
  which consists of four molecules per unit cell and 
   3/4-filled band (quarter filling of hole)  with 
     eight transfer energies.
     \cite{Mori} 
For ambient pressure, 
   \eti shows a metal-inslator transition 
 at $T_{MI}=135$ K 
     followed  by the charge ordering. 
\cite{Takano} 
Under hydrostatic pressure, 
   $T_{MI}$ decreases and vanishes  where,   
    at 20 kbar, 
\cite{Tajima_2000} 
    the resistivity becomes almost temperature-independent 
  from 300 K to 1.5 K. 
The carrier density decreases from $10^{21}$ cm$^{-3}$ (at 300 K) 
  to $10^{15}$ cm$^{-3}$ (at 1 K) 
   while the carrier mobility is interpreted to move 
    from $10^{-1}$ cm$^{2}$/(V$\cdot$s) (at 300 K) to $10^5$ cm$^{2}$/(V$\cdot$s) (at 1 K). 
It has been maintained that 
 such a temperature-dependence of the carrier indicates 
  a  narrow gap semiconducting (NGS) state 
   with the gap $E_g\sim 1$meV\cite{Tajima_2000}. 
  The NGS state
  is also obtained  at 10 kbar, 
  \cite{Tajima_2002} 
   under the uniaxial strain\cite{Maesato} 
    along $a$-direction. 
 
The  band calculation of \eti shows 
 a zero gap semiconducting (ZGS) state 
 \cite{Kobayashi_JPSJ,Kobayashi_JPSJ2}  
 instead of the NGS state
  when the uniaxial pressure along $a$-axis, $p_a$, becomes large.  
The Fermi surface of \eti salt at ambient pressure 
  shows a semi-metallic state with the hole and the electron pocket,
 \cite{Tajima_2000} 
    and the band spectrum has the degeneracy 
      at the bottom of the electron pocket
        with an incommensurate wave number $\pm\blet{k}^0$.
        \cite{zgsjpsj} 
Since $\blet{k}^0$ is different from the location corresponding to 
    the symmetry of the Hamiltonian, 
     such a degeneracy is called as  accidental one 
       according to Herring\cite{Herring}. 
With increasing  $p_a$, the hole and electron pockets are reduced
\cite{Kondo_France} 
  and diminish, for $p_a>3$ kbar,
    with the resultant band dispersion, 
   showing  Dirac cones   centered on $\pm\blet{k}^0$. 
\cite{zgsjpsj} 
Such a band dispersion and the corresponding density of states 
  per spin and per volume are given respectively by 
\begin{align}
 \xi_{\mib{k}} &=  
 \pm \hbar v |\blet{k}- \blet{k}^0|, 
             \nonumber\\
  D(\vep) &=\dfrac{|\vep|}{\pi \hbar^2v^2} ,
\label{lineardos}
\end{align}
 close to the Fermi energy $\vep_F$,  
   where $v$ is the average of the velocity and $\vep$ is the energy measured from $\vep_F$.  

In terms of eq. (\ref{lineardos}), the carrier density $n$ 
  of the ZGS state at low temperatures is calculated as  
\begin{align}
n=2\int_{-\infty}^{\infty}D(\vep)f(\vep)d\vep=
\dfrac{\pi^2(k_BT)^2}{3}\dfrac{1}{\pi \hbar^2 v^2},
\label{n}
\end{align}
 where $f(\vep)=(e^{\vep/k_B T}+1)^{-1}$ and $T$ is the temperature. 
 Equation (\ref{n})  is consistent with that of the experiment.\cite{Tajima_2000,Tajima_2002} 
However, it is unclear 
 if the increase of the carrier density is canceled by 
   the decrease of the carrier mobility 
   when the conductivity keeps  constant with increasing  temperature. 
  In the present paper,  the conductivity as the funcion of $T$ and $\tau$ 
  is examined  for the ZGS state  of $\alpha$-(ET)$_2$I$_3$.  
 The calculation is performed by using  the Green's function 
   without the vertex corrections 
   where  the life time $\tau$  in the self energy  is treated 
      as a parameter.

\begin{figure}[tbp]
\begin{center}
\includegraphics[width=8.0cm]{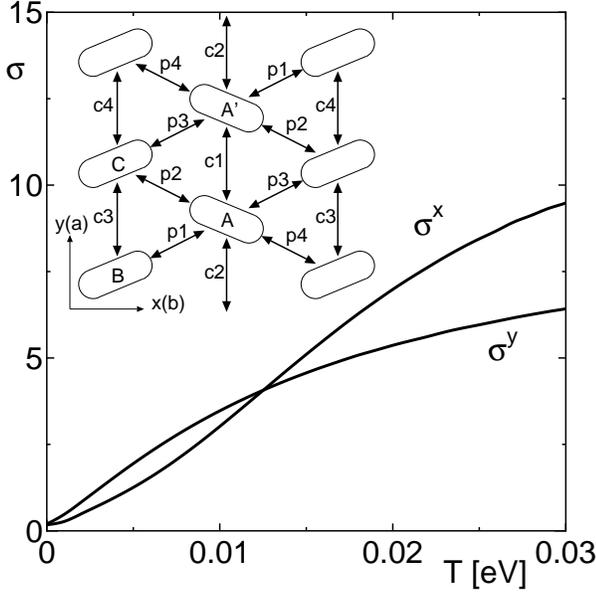}
\caption{Temperature dependence of $\sigma^{x}$ and $\sigma^{y}$ for ZGS state at 
         $p_a=10$ (kbar), where $1/\tau=0.001$ (eV). 
         The inset shows the structure of \eti on the conducting plane.}
\label{tlinear}
\end{center}
\end{figure}

In the  system with multi-sites, 
  we consider explicitly  $H_0$ for the transfer energy  which is crucial to 
 obtain the ZGS state,  but treat  implicitly  $H_{int}$ for 
  the interaction  which gives rise to  the life time.  
 The coupling to the external field, $H_{ext}$ is also  added 
     to calculate the conductivity. 
  The total Hamiltonian is written as 
 $ H =  H_0 +  H_{int} + H_{ext}$, where 
  
\begin{align}
H_0 &= 
     \sum_{n.n.}t_{i\alpha:j\beta}a^{\dag}_{i\alpha}a_{j\beta} , 
\label{h00}\\
H_{ext} &= 
 -e\sum_{i\alpha}\blet{r}_i\cdot\blet{E}a^{\dag}_{i\alpha}a_{i\alpha}
=-A
 ,
\label{ht}
\end{align}
and $i,j(=1,\cdots,N_L)$ denotes indices for the unit cell, 
forming a square lattice. 
The indices $\alpha,\beta$ denote the site in the unit cell.
The quantity $t_{i\alpha:j\beta}$ is the transfer energy for the
 nearest neighbor (n.n.) sites between 
 $(i,\alpha)$  and $(j,\beta)$, and 
$a_{i\alpha}$ is the annihilation operator of the conduction electron at $(i,\alpha)$ site. 
In eq. (\ref{ht}), $\blet{r}_i$ is the position vector of the unit cell coupled with 
the electric field $\blet{E}$ where $e$ is the electric charge. 
Since the equation of motion for $A$ is written as 
$\dot{A}=(i/\hbar)[H,A]=\blet{J}\cdot\blet{E}$, 
the current density $J_{\mu}$ ($\mu$ 
   being the direction of $\blet{J}$) is given as 
\begin{align}
J_{\mu}&=\dfrac{e}{\hbar}\sum^{}_{\mib{k}\alpha\beta}
\dfrac{\partial\vep_{\alpha\beta}(\blet{k})}{\partial k_{\mu}}
a^{\dag}_{\mib{k}\alpha}a_{\mib{k}\beta}=
\dfrac{e}{\hbar}\sum_{\mib{k}\gamma\gamma'}
v^{\mu}_{\gamma\gamma'}(\blet{k})
c^{\dag}_{\mib{k}\gamma}c_{\mib{k}\gamma'},
\label{j}
\end{align}
where 
\begin{align}
\vep_{\alpha\beta}(\blet{k})&=
\dfrac{1}{V}\sum_{n.n.}t_{i\alpha:j\beta}
e^{i\mib{k}\cdot(\mib{r}_i-\mib{r}_j)},
\label{hm}
\\
v^{\mu}_{\gamma\gamma'}(\blet{k})&=
\sum^{4}_{\alpha\beta=1}
d^{*}_{\alpha\gamma}(\blet{k})
\dfrac{\partial \vep_{\alpha\beta}(\blet{k})}{\partial k_{\mu}}
d_{\beta\gamma'}(\blet{k}),
\label{vmu}
\end{align}
and $V$ is the area of a system. 
The directions $x$ and $y$ correspond to $a$ and $b$
     in the Mori's notation, respectively 
              (inset of Fig. \ref{tlinear}).
              \cite{Mori} 
The quantities $a_{\mib{k}\alpha}$ and $c_{\mib{k}\gamma}$ denote 
   the Fourier transform, 
$a_{\mib{k}\alpha}= 
V^{-1/2}
\sum_{i=1}^{N_L}
a_{i\alpha}e^{-i\mib{k}\cdot\mib{r}_i},\ 
c_{\mib{k}\gamma}=\sum_{\alpha=1}^{4}
d^{*}_{\alpha\gamma}(\blet{k})a_{\mib{k}\alpha}$, 
   where the band index $\gamma ( = 1, \cdots,4) $.
The quantity $d_{\alpha\gamma}(\blet{k})$ is the component of the eigenvector 
  obtained  by 
\begin{align}
\dsum{\beta=1}{4}
(\vep_{\alpha\beta}(\blet{k})-\vep_F)d_{\beta\gamma}(\blet{k})=
\xi_{\mib{k}\gamma}
d_{\alpha\gamma}(\blet{k}),
\label{eigensystem}
\end{align}
where $\xi_{\mib{k}\gamma}$ 
 ($\xi_{\mib{k} 1} >  \xi_{\mib{k} 2} > \xi_{\mib{k} 3} > \xi_{\mib{k} 4}$) 
is the eigenvalue of the multi-band 
    with the state $(\blet{k},\gamma)$. 
The matrix element  
  $\vep_{\alpha\beta}(\blet{k})$ is given by  eq. (\ref{hm}), 
    which depends on    eight transfer energies, $t_A$,  
        as shown in the inset of Fig. \ref{tlinear}. 
  The indices 
     $A=c1,c2,c3,c4,p1,p2,p3,p4$  correspond  to 
         $a2,a3,a1,a1,b2,b1,b4,b3$ , respectively, in the definition 
          by   Mori \etal \cite{Mori}
 Since the ZGS state is 3/4-filling, 
    the band of $\xi_{\mib{k}1}$ is empty and the others 
are filled at $T=0$. 

In terms of the linear response theory,  
 the electric conductivity $\sigma^{\mu\nu}$, $(\mu, \nu = x, y )$
   is expressed as, 
\begin{align}
\sigma^{\mu\nu}=
i[(Q^{R}_{\mu\nu}(\omega)-Q^{R}_{\mu\nu}(0))/\omega]_{\omega\rightarrow0}\; . \;
\label{kubo}
\end{align}
The retarded Green's function
  $Q^{R}_{\mu\nu}(\omega)$ is obtained from 
 the  analytical continuation of  $Q_{\mu\nu}(i \omega_l)$ 
  where $Q_{\mu\nu}(i \omega_l)$  
   is the  two-body Green's function (T$_i$ is the ordering operator) of
     imaginary time $t_i$, given by   
\begin{align}
Q_{\mu\nu}(i\omega_l)=-\frac{1}{V}\int_{0}^{\beta}
\langle\mathrm{T}_i J_{\mu}(t_i)J_{\nu}\rangle 
e^{i\omega_l t_i} {\rm d} t_i, 
\label{q}
\end{align}
  where $\omega_l=2\pi k_BTl$ with $l$ being an integer. 

Equation (\ref{q}) is calculated by adopting one-loop approximation, 
    i.e. without vertex correction. 
We treat the effect of $H_{int}$ as a life time, $\tau$, 
  in the Green's function,
  \cite{Abrikosov} 
\begin{align}
G_{\gamma\gamma'}(\blet{k},i\vep_m)&=
-\int^{\beta}_{0}\langle\mathrm{T}_i
c_{\mib{k}\gamma}(t_i)c^{\dag}_{\mib{k}\gamma'}(0)
\rangle
e^{i\vep_{m}t_i} {\rm d}t_i
\nonumber\\
&\simeq
        (i\vep_{m}-\xi_{\mib{k}\gamma}
        -\Sigma(i\vep_m))^{-1}\delta_{\gamma\gamma'} \;, \; 
\label{G}\\
\Sigma(i\vep_m)&=
        -i\hbar(\mathrm{sgn}(\vep_m)/2\tau),
\label{self}
\end{align}
   where $\vep_{m}=(2m+1)\pi k_BT$ with $m$ being an integer. 
The life time, $\tau$, is naively taken as a parameter where $\tau$ may come from the processes 
   such as the scattering by electron-phonon interaction, 
   impurity and fluctuation. 
Within the above approximation, eq. (\ref{q}) is expressed as 
\begin{align}
Q_{\mu\nu}(i\omega_l)&=\frac{e^2 k_BT}{\hbar^2 V}
   \sum_{\mib{k}\gamma\gamma'm}
   v^{\mu}_{\gamma\gamma'}(\blet{k})
   v^{\nu}_{\gamma'\gamma}(\blet{k})
\nonumber\\
   &\times 
   G_{\gamma'\gamma'}(\blet{k},i(\varepsilon_{m}+\omega_l))
   G_{\gamma\gamma}(\blet{k},i\varepsilon_{m})\nonumber\\
&=
   \dfrac{e^2}{\pi \hbar^2 V}\sum_{\mib{k}\gamma\gamma'}
  v^{\mu}_{\gamma\gamma'}(\mib{k})
  v^{\nu}_{\gamma'\gamma}(\mib{k})
  \int_{-\infty}^{\infty}d\vep f(\vep)\nonumber\\
  &\times
  \left[\dfrac{1}{\vep+i\omega_l-\xi_{\mib{k}\gamma'}+(i\hbar/2\tau)}
  \dfrac{(\hbar/2\tau)}{(\vep-\xi_{\mib{k}\gamma})^2+(\hbar/2\tau)^2}\right.\nonumber\\
&+
  \left.\dfrac{1}{\vep-i\omega_l-\xi_{\mib{k}\gamma}-(i\hbar/2\tau)}
  \dfrac{(\hbar/2\tau)}{(\vep-\xi_{\mib{k}\gamma'})^2+(\hbar/2\tau)^2}\right]. 
\label{calq}
\end{align}
  After the analytical continuation
  (i.e., $i \omega_l \rightarrow \hbar\omega + i 0 )$ , 
  the real part of $\sigma^{\mu\nu}$ ($\mu=\nu=x,y$) per spin is calculated as, 
\begin{align}
\sigma^{\mu}&=\mathrm{Re}(\sigma^{\mu\mu})=\sum_{\gamma\gamma'}
\sigma^{\mu}_{\gamma\gamma'},\nonumber\\
\sigma^{\mu}_{\gamma\gamma'}&=\frac{e^2}{\pi \hbar V}
\sum_{\mib{k}}
|v^{\mu}_{\gamma\gamma'}(\blet{k})|^2
\int_{-\infty}^{\infty}d\vep \left(-\dfrac{\partial f(\vep)}{\partial \vep}\right)
\nonumber\\
&\times
\frac{(\hbar/2\tau)}
{(\vep-\xi_{\mib{k}\gamma'})^2
+(\hbar/2\tau)^2}
\frac{(\hbar/2\tau)}
{(\vep-\xi_{\mib{k}\gamma})^2
+(\hbar/2\tau)^2}.
\label{sigmamumu}
\end{align}

Based on the experimental data, 
  we use the interpolation formula for the transfer energy, where 
  $t_{A}$ (eV)  at $p_a$ (kbar) is given by , 
  \cite{Kobayashi_JPSJ,Kondo_France} 
\begin{align}
t_A(p_a)= t_A(1+K_A p_a) . 
\label{transfer_P}
\end{align}
The units of the energy and the 
  pressure are taken as eV and kbar, respectively. 
In eq. (\ref{transfer_P}), 
  $t_{c1}=0.048$, $t_{c2}=-0.020$, $t_{c3}=-0.028$, $t_{c4}=-0.028$, 
  $t_{p1}=0.140$, $t_{p2}=0.123$, $t_{p3}=-0.025$, $t_{p4}=-0.062$, 
  $K_{c1}=0.167$, $K_{c2}=-0.025$, $K_{c3}=0.089$, $K_{c4}=0.089$, 
  $K_{p1}=0.011$, $K_{p2}=0.000$, $K_{p3}=0.000$ and $K_{p4}=0.032$. 
The ZGS state is obtained for $p_a>3$\cite{zgsjpsj}. 
We calculate $\sigma$ as the function of $T$ and $\tau$ where 
  the area of an unit cell $V/N_L$, 
  the electric charge $e$, Plank constant $\hbar$ and Boltzmann constant $k_B$ 
  are set unity. 

\begin{figure}[tbp]
\begin{center}
\includegraphics[width=8.0cm]{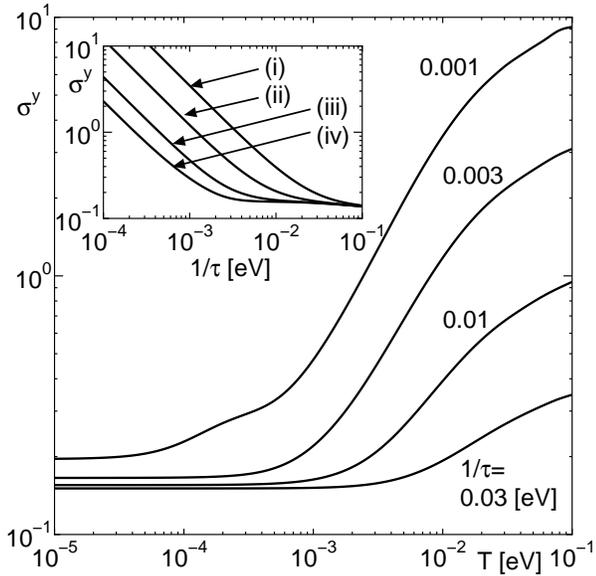}
\caption{Temperature dependence of $\sigma^{y}$ at 
         $p_a=10$ where the inset shows the $1/\tau$ dependence of $\sigma^{y}$ 
         at the temperature $T=0.01$(i),$0.003$(ii),$0.001$(iii), and $0.0003$(iv) 
         respectively.
}
\label{tlog}
\end{center}
\end{figure} 
Figure \ref{tlinear} depicts the temperature dependence of $\sigma$ for 
  $1/\tau=0.001$ and $p_a=10$ (ZGS state). 
The difference between $\sigma^x$ and $\sigma^y$ comes from the anisotropy of $t_A$, 
  where $\sigma^x$ is larger(smaller) than $\sigma^y$ for $T\gtrsim0.14$ ($T\lesssim0.14$) 
  in the present choice of parameters. 
In Fig. \ref{tlinear}, one finds $\sigma^{\mu}\propto T$ at low temperatures in contrast to 
  the conventional conductivity 
  given by the Drude's formula (i.e. $\sigma=ne^2\tau/m$ with $m$ being the electron mass). 
Thus it is incorrect that  $\sigma^{\mu}$ in the ZGS state 
  is  propotional to the carrier number with $n (\propto T^2)$. 
The decrease of $\sigma^{\mu}$ at low temperatures comes from 
  the density of states $D(\vep)$, which is given by eq. (\ref{lineardos}) near the Fermi energy. 

Figure \ref{tlog} shows the temperature dependence of $\sigma^{y}$ for several choices of 
  life time $1/\tau$.
In the high temperature region, 
  the conductivity increases monotonically up to $T=0.1$, and is large for small $1/\tau$. 
In the low temperature region, $\sigma^y$ is independent of the temperature 
This region is wider when $1/\tau$ is larger. 
In the inset  the $1/\tau$ dependence of   $\sigma^y$ is shown
  for  $T=0.01, 0.003, 0.001$ and $0.0003$. 
The  $\sigma^y$ is flat for  large $1/\tau$ 
 and the gradient of $\sigma^y$ is about $1$
 for small $1/\tau$ indicating that 
  $\sigma^y$ is propotional to $\tau$. 
The above result is different from the conventional metallic state 
   in which the conductivity is always given by $\sigma=ne^2\tau/m$. 
The region in which the conductivity is independent of the life time is wide when 
  the temperature is low. 

\begin{figure}[tbp]
\begin{center}
\includegraphics[width=8.0cm]{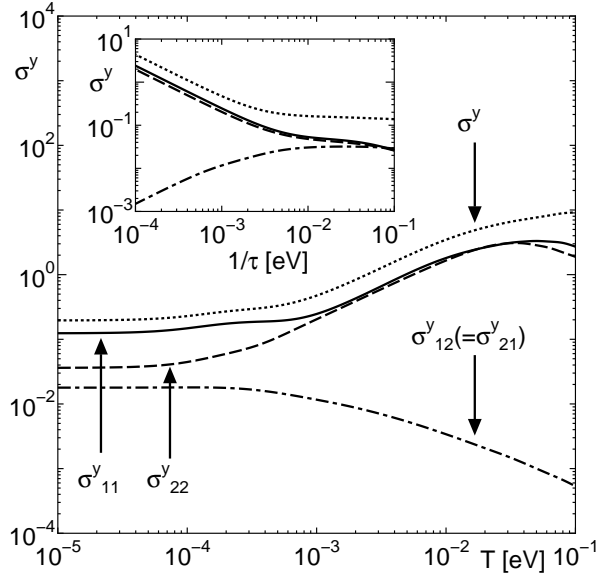}
\caption{
         Main figure: temperature dependence of $\sigma^{y}$ (dotted line), 
         $\sigma^{y}_{11}$ (solid line), 
         $\sigma^{y}_{22}$(dashed line) and 
         $\sigma^{y}_{12}(=\sigma^{y}_{21})$ (dot-dashed line) at 
         $p_a=10$ and $1/\tau=0.001$. 
         Inset: the corresponding quantitiy as the function of 
          $1/\tau$ at $ p_a = 10$ and $T = 0.001$. 
}
\label{taulog}
\end{center}
\end{figure} 
In Fig. \ref{taulog}, the temperature dependence of the components of 
  $\sigma^{y}_{\gamma\gamma'}$ with $(\gamma,\gamma')=(1,1),(1,2),(2,2)$ and $\sigma^{y}$ 
  are shown for $p_a=10$ and $1/\tau=0.001$. 
The corresponding conductivity as the function of $1/\tau$ ($T=0.001$) 
  is also depicted in the inset. 
The diagonal components of $\sigma^{y}_{11}$ and $\sigma^{y}_{22}$ 
  mainly contribute to $\sigma^{y}$, whereas the off-diagonal component 
  $\sigma^{y}_{12}(=\sigma^{y}_{21})$ contributes 
 only at the low temperatures. 
The fact that $\sigma^{y}_{12}$ decreases monotonically in contrast to 
  $\sigma^{y}_{11}$ and $\sigma^{y}_{22}$ in the region of high temperature 
  is understood as follows. 
The carriers excited thermally can participate in $\sigma^y_{11}$ 
 and $\sigma^y_{22}$ 
  while only carriers close to Fermi surface give rise to $\sigma^y_{12}$ and $\sigma^y_{21}$. 
For the $1/\tau$ dependence, 
  $\sigma^{y}_{11}$ and $\sigma^{y}_{22}$ are propotional to $\tau$, and 
  $\sigma^{y}_{12}$ decreases for small $1/\tau$. 
  Such a behavior of $\sigma^y_{12}$ 
  is also understood by noting that 
   the corresponding height in  the r.h.s. of eq.(\ref{sigmamumu}) 
  at $\vep = 0$ becomes large for large $1/T$ and large $1/\tau$. 
Thus $\sigma^{y}_{12}$ can contribute in the dirty case of $T\tau\ll1$. 
Compared with these four components of $\sigma^{y}_{11},\sigma^{y}_{22},\sigma^{y}_{12}$, 
  and $\sigma^{y}_{22}$, 
  the other twelve components of $\sigma^{y}_{\gamma\gamma'}$ are negligibly small. 


The behavior of the conductivity in Fig. \ref{tlog} can be explained from the property of 
  the density of states in eq. (\ref{sigmamumu}). 
Since $D(\omega)=\sum_{\mib{k}}\delta(\omega-\xi_{\mib{k}})=(|\omega|/2\pi v^2)$ with 
  $\pm\xi_{\mib{k}}=v\sqrt{k_x^2+k_y^2}$ ($v$ is an averaged velocity), 
  eq. (\ref{sigmamumu}) 
   with only the diagonal component of $(\gamma,\gamma')=(1,1)$ and $(2,2)$ 
  is calculated as ($x=\beta\vep$), 
\begin{align}
\sigma&\sim
\dfrac{v^2}{\pi}
\int_{-\infty}^{\infty}d\omega
d\vep
D(\omega)\left(-\dfrac{\partial f(\vep)}{\partial\vep}\right)
\left[\frac{(1/2\tau)}
{(\vep-\omega)^2
+(1/2\tau)^2}\right]^2
\nonumber\\
&=\dfrac{1}{4\pi^2}
\int_{-\infty}^{\infty}dx
\dfrac{1+(2T\tau |x|)\tan^{-1}(2T\tau |x|)}{\cosh x+1}.
%
\label{sigmar}
\end{align}
When $T\gg 1/\tau$, one can replace $\tan^{-1}(2T\tau|x|)$ by $\pi/2$ and estimate 
  eq. (\ref{sigmar}) as 
\begin{align}
\sigma&\simeq
(\log2/\pi) T\tau\simeq 0.220 T\tau.
\label{propotional}
\end{align}
Thus $\sigma$ is propotional to the temperature and the life time for $T\gg 1/\tau$. 
The results of eq. (\ref{propotional}) is also derived by using 
  Boltzmann's equation where the conductivity, $\sigma_B$, is expressed as 
\begin{align}
\sigma_B=\dint{-\infty}{\infty} v^2 \tau D(\vep)
\left(-\dfrac{\partial f(\vep)}{\partial \vep}\right)d\vep. 
\label{boltzmann}
\end{align}
Actually eq. (\ref{sigmar}) becomes equal to eq. (\ref{boltzmann}) for $|\vep|\gg1/\tau$, 
  indicating that the case $T\gg1/\tau$ corresponds to the classical limit. 
When $T\ll 1/\tau$, eq. (\ref{sigmar}) is calculated as 
\begin{align}
\sigma\simeq 1/(2\pi^2)=0.051,
\label{const}
\end{align}
which shows $\sigma$ being indepnedent of both temperature and life time. 
In  both cases of $T\gg 1/\tau$ and $T\ll 1/\tau$, the conductivity does not 
  depend on the velocity $v$. 
\begin{figure}[tbp]
\begin{center}
\includegraphics[width=7.0cm]{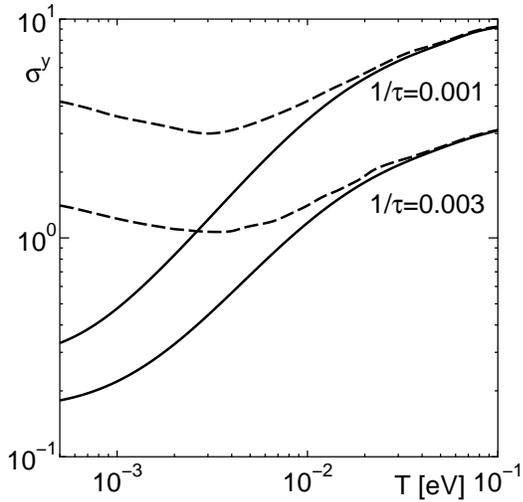}
\caption{
Conductivities by Born approximation (dashed line)
, which are compared with those of  Fig. 2 (solid line).
}
\label{talpha}
\end{center}
\end{figure} 
These two regions of $\sigma\sim const.$ and $\sigma\sim T\tau$ 
  are seen in Fig. \ref{tlog}.
  The boundary between two the region of  $\sigma\sim const.$ and 
   that of $\sigma\sim T\tau$ 
   is obtained as 
  $1/\tau\simeq 0.7T$ at $p_a=10$ and $1/\tau\simeq 0.8T$ at $p_a=6$. 
Equaions (\ref{propotional}) and (\ref{const}) exhibit 
  the universal coefficient of $T\tau$ and the universal constant, 
     respectively.
These values multiplied by 2 are compared with those of Fig. \ref{tlog}, 
   because eqs. (\ref{propotional}) and (\ref{const}) are calculated 
    in a single Dirac cone.
For $0.001<T<0.005$, 
    the coefficient of $T\tau$ in Fig. \ref{tlog} corresponds  well 
      to the universal coefficient.
At low temperature, however, the conductivity in Fig. \ref{tlog} 
  is somewhat larger than the universal constant, 
    because the former contains  the contribution from  
     the off-diagonal components,   $(\gamma,\gamma')=(1,2),(2,1)$.

Now we discuss the relevance of the  ZGS state 
to the conductivity of \eti, which is  almost  independent of temperature. 
In terms of the mobility and the carrier, \cite{Tajima_2002} 
 the bulk conductivity is estimated  as 
$\sigma \simeq 3.2 \times 10^{2} \Omega^{-1}$ cm$^{-1}$. 
Then one obtains   $1/\tau \simeq  $ 1K  at room temparature
  by using naively   the formula  
   $\sigma = n e^2 \tau /m$ 
    with the bare mass $m$.
   Since   the present result of the temperature dependence 
    for $T > 10 K$    (i.e., $T \tau \gg 1$ )
         is not consistent with that of the experiment,  
    we calculate  the life  time of  the impurity scattering 
     by taking account of  the ZGS state. 
 In terms of the self-consistent Born approximation 
 \cite{Abrikosov}, the self-energy in eq.~(\ref{self}) is replaced by 
\begin{align}
\Sigma(i\vep_m) =
        -i\hbar \vep_m /(2\tau W_D) , 
\label{self_Born}
\end{align}
for $\vert \vep_m \vert < W_D$, where $W_D$ 
 ($\sim 0.01$ at $p_a$ = 10 kbar) denotes the band width  of the Dirac cone.
Note that the  frequency dependence similar to eq. ({\ref{self_Born}}) 
has been obtained for the calculation in the graphite at $T=0$. 
\cite{Gonzalez,Shon_Ando} 
 Thus   we obtain the  conductivity in which the temperature dependence is reduced as shown  in Fig.~\ref{talpha}
   indicating a tendency being  consistent with the experiment. 
 However there remains a problem of the magnitude of $\sigma$
  which comes from  the stacking of the two-dimensional sheet  of the ET salt. 
From the balk conductivity  and   the lattice constant 
 of $c$-axis (as $1.7 \times 10^{-7}$ cm),
  \cite{Kondo_France}
 the conductivity  per one sheet and spin,  is estimated as 
 0.11  in  the unit of 
  $e^2/\hbar (= 2.44 \times 10^{-5} \Omega^{-1}$),
 which is utilized    for the conductivity in the present calaculation. 
    This magnitude is 
     is much smaller than the numerical result  of Fig.~\ref{talpha}
     although the origin of  such a discrepancy is not clear for the 
     moment.  
  
Here, we show  the validity of the  extrapolation formula
   eq.~(\ref{transfer_P})  even for 10 kbar by comparing the coeffients of 
    the density of states in eq. (\ref{lineardos}).
 The experimental value of the   carrier density at $T$  is given by 
     $n \simeq 10^{15} T^2$  cm$^{-3}$ and then 
        the density per sheet is given by 
           $n \simeq 1.7 \times 10^{8} T^2$  cm$^{-2}$.
  By comparing the coefficient  with that of eq.~(\ref{n}) 
   ( lattice constants are   $a = 9.2 \times 10^{-8}$ cm and 
   $b = 10.8 \times 10^{-8}$ cm ), \cite{Kondo_France} 
       the coefficients of $\vep$ in eq.~(\ref{lineardos}) 
           is estimated as $8 \times 10$  eV$^{-2}$, 
           which corresponds well 
            to  that the present calculation
              $\sim$  6 $\times$ 10 eV$^{-2}$.  

  Finally we note that the three dimensional effect on the ZGS state.
Since the small interplane hopping is expected in the \eti 
salt from the first principle calculation,
\cite{Kino_Miyazaki} 
the metallic state with small Fermi surface appears.\cite{Mikitik} 
Then the property of the ZGS state could be found at temperature larger than the interplane hopping.

\section*{Acknowledgements}
The authors are thankful to  K. Kajita, N. Tajima, H. Kontani, H. Yoshioka and H. Fukuyama for useful discussions. 
The present work has been financially supported 
by a Grant-in-Aid for Scientific Research on Priority Areas of Molecular Conductors 
(No. 15073103) from the Ministry of Education, Culture, Sports, Science and Technology, Japan.


\end{document}